\documentclass[superscriptaddress,showpacs,amsmath,amssymb,twocolumn,nofootinbib]{revtex4-1}
\usepackage{amsmath, amsthm, amssymb, bm, braket}
\usepackage[dvips]{graphicx}
\usepackage{color}
\usepackage[normalem]{ulem}
\newcommand{\bi}[1]{\ensuremath{\boldsymbol{#1}}}

\newcommand{\oprt}[1]{\ensuremath{\hat{\mathcal{#1}}}}
\newcommand{\abs}[1]{\ensuremath{\left| #1 \right|}}

\begin{document}
\title{Discerning nuclear pairing properties from magnetic dipole excitation}
\author{Tomohiro Oishi}
\email[E-mail: ]{toishi@phy.hr}
\affiliation{Department of Physics, Faculty of Science, University of Zagreb, Bijeni\v{c}ka c. 32, 10000 Zagreb, Croatia}
\author{Goran Kru\v{z}i\'{c}}
\email[E-mail: ]{goran.kruzic@ericsson.com}
\affiliation{Research department, Ericsson - Nikola Tesla, Krapinska 45, 10000, Zagreb, Croatia}
\affiliation{Department of Physics, Faculty of Science, University of Zagreb, Bijeni\v{c}ka c. 32, 10000 Zagreb, Croatia}
\author{Nils Paar}
\email[E-mail: ]{npaar@phy.hr}
\affiliation{Department of Physics, Faculty of Science, University of Zagreb, Bijeni\v{c}ka c. 32, 10000 Zagreb, Croatia}
\def \beq{\begin{equation}}
\def \eeq{\end{equation}}
\def \beqa{\begin{eqnarray}}
\def \eeqa{\end{eqnarray}}
\def \bir{\bi{r}}
\def \ubir{\bar{\bi{r}}}
\def \bip{\bi{p}}
\def \ubip{\bar{\bi{r}}}
\def \adel{\tilde{l}} 
\def \colourb{\color{black}}
\def \colourr{\color{black}}
\def \colourbb{\color{black}}
\begin{abstract}
Pairing correlation of Cooper pair is a fundamental property of multi-fermion interacting systems. For nucleons, two modes of the Cooper-pair coupling may exist, namely of $S_{12}=0$ with $L_{12}=0$ (spin-singlet s-wave) and $S_{12}=1$ with $L_{12}=1$ (spin-triplet p-wave).
In nuclear physics, it has been an open question whether the spin-singlet or spin-triplet coupling is dominant, as well as how to measure their role. 
We investigate a relation between the magnetic-dipole (M1) excitation of nuclei and the pairing modes within the framework of relativistic nuclear energy-density functional (RNEDF).
The pairing correlations are taken into account by the relativistic Hartree-Bogoliubov (RHB) model in the ground state, and the relativistic quasi-particle random-phase approximation (RQRPA) is employed to describe M1 transitions. We have shown that M1 excitation properties display a sensitivity on the pairing model involved in the calculations. The systematic evaluation of M1 transitions together with the accurate experimental data enables us to discern the pairing properties in finite nuclei.
\end{abstract}
\maketitle

\section{Introduction} \label{intro}
Pairing correlation is a fundamental property of multi-fermion interacting systems.
In the original Bardeen-Cooper-Schrieffer (BCS) picture \cite{57Bardeen}, the pairing correlation is attributed to the phonon-mediated effective interaction, where the two electrons are coupled to have the spin-singlet (SS) Cooper pair. Its total spin is $S_{12}=0$ with $L_{12}=0$ (s-wave).
In the superfluidity of the Helium-3 atoms \cite{1972Osheroff_01,1972Osheroff_02,1972Leggett}, on the other hand, two fermionic atoms are expected to couple into the spin-triplet (ST) pair with $S_{12}=1$ and $L_{12}=1$ (p-wave), due to the spin-fluctuation effect. 
These pairing correlations lead to the emergence of the pairing gap as well as the Bose-Einstein condensate.

In atomic nuclei, the nuclear Cooper pair has been interpreted as a consequence of the {\colourb residual interaction \cite{03Dean_rev,05BB,13BZ}. }
The pairing interaction plays an essential role to determine the binding energy, which shows so-called odd-even staggering behaviour relevant to the pairing gap energy.
{\colourb 
For its evaluation, theoretical models of the attractive force in the SS-pairing channel have been utilized \cite{96Jacek,04Suzu,2005Sand,08Hagi,2008Bulgac,2013Pastore}.
On the other side, the attractive ST-pairing has been also considered in some calculations.
For example, an attractive force in the ST channel of the valence neutrons has been suggested in the dilute density, i.e. around the surface of nuclei \cite{1968Tamagaki,1970Tamagaki}.
In several calculations, the pairing model with the attractive force in both the SS and ST channels has been employed \cite{1980Gogny,1991Berger,88Suzuki_COSM,03Myo,10Myo,2014Oishi}.
}
Depending on the chosen model, the nuclear Cooper pair is expected to have the $S_{12}=0$ and/or $S_{12}=1$ components in the valence orbit(s). One can often find that different pairing models somehow equivalently reproduce the empirical energies, as long as they are adjusted consistently to the reference data. Namely, the finite ambiguities remain in the pairing modes in medium.

In our recent studies \cite{2019OP,2020Oishi}, it has been shown that the magnetic-dipole (M1) excitation is closely connected with the intrinsic components of the coupled spin $S_{12}$. 
There, if the SS pair is dominant, the M1-excitation strength as well as its summation value is suppressed.
Its excitation energy also depends on the model to describe the effective-pairing force, which can support the $S_{12}=0$ or $S_{12}=1$ mode of pairing.
Therefore, by evaluating the M1 excitation, it may provide a suitable way to resolve the pairing-mode ambiguities.
Note also that, 
for the description of M1 mode, the spin-orbit (SO) splitting and the M1-residual interactions essentially contribute \cite{1985Rich,1990Rich,2010Heyde_M1_Rev,2008Pietralla_Rev,2011Fujita_M1_GT,2006Speth}.
For example, in studies based on the Skyrme energy-density functional, the interplay between the SO splittings and the residual interactions has been intensely studied \cite{2009Vesely,2010Nest,2010Nest_2,2019Tselyaev,2020Speth}.
In ref. \cite{2020Oishi,2020Kruzic}, we discussed the same topic but in the relativistic nuclear energy-density functional (RNEDF) framework.

In this work, we investigate the isovector (IV) M1 excitation focusing on its pairing-mode sensitivity, i.e. the role of the SS and ST pairing channels. 
Our theoretical framework is based on the RNEDF theory \cite{1974Walecka,1977Boguta,1989Reinhard,2005Vret,2006Meng,2016Ebran_PRC}.
From the scalar and vector couplings within the Dirac-Lorentz structure of the formalism, the RNEDF theory naturally describes the SO splittings, which are essential in modeling the M1 excitation.
The relativistic Hartree-Bogoliubov (RHB) method is used to determine the ground state (GS), 
while the relativistic quasi-particle random-phase approximation (RQRPA) is employed to describe the M1-excitation properties, as introduced in ref. \cite{2020Kruzic,2020Oishi}. 
More details about the theory framework are given in ref. \cite{2020Oishi,2020Kruzic,2003Paar,2008Niksic,2014Niksic}. 

In the next section, the theoretical and numerical details are presented. 
Our main results and discussions are presented in sec. \ref{Sec:III}. 
Finally in sec. \ref{Sec:summary}, we summarize this work. 
In this work, the spherical symmetry is assumed, and we employ the CGS-Gauss system of units.

\section{Formalism and setting} \label{sec:form}
Parameters in this work are fixed as the same in ref. \cite{2020Oishi}, except when noticed. 
Namely, for the particle-hole channel, we employ the density-dependent point-coupling interaction with the DD-PC1 parameterization \cite{2020Kruzic,2020Oishi,2014Niksic,2008Niksic}. 
In addition, for the RQRPA description of M1 mode, the isovector-psudovector (IV-PV) coupling term is employed to describe the RQRPA residual interaction in the particle-hole channel \cite{2020Oishi,2020Kruzic}. 
The total Lagrangian density reads 
\beq
 \mathcal{L} = \mathcal{L}_{\rm DD\mathchar"712D PC1} + \mathcal{L}_{\rm IV \mathchar"712D PV} - e\bar{\psi} \gamma^{\mu} A_{\mu} \frac{1-\hat{\tau}_3}{2} \psi(x).
\eeq
The DD-PC1 Lagrangian includes the isoscalar-scalar, isoscalar-vector, and isovector-vector channels, where their derivative terms are also employed for a quantitative description of the nuclear density distribution \cite{2008Niksic}. 
The IV-PV term, on the other side, gives a finite contribution only in the RQRPA solution of the $1^+$ states \cite{2020Oishi,2020Kruzic}. 
In addition, the coupling of protons to the electromagnetic field is taken into account. 
For the particle-particle channel, we employ the finite-range Gaussian pairing potential. 
Note that the proton-neutron pairing is omitted in this work. 

We investigate the IV-M1 excitation up to the one-body-operator QRPA level. 
{\colourbb By using the isospin $\hat{\tau}_3=1~(-1)$ for protons (neutrons) and 
the nuclear magneton $\mu_{\rm N}$, 
the IV-M1 operator reads 
\beq
  \oprt{Q}_{\nu}^{\rm (IV)} = \sqrt{\frac{3}{4\pi}} \mu_{\rm N} \sum_{i \in A} \left[ g^{\rm (IV)}_l\hat{l}_{\nu}(i)  +g^{\rm (IV)}_s\hat{s}_{\nu}(i) \right] \hat{\tau}_3(i), \label{eq:e6ag32r}
\eeq
in the SP format including the spin $\hat{s}_{\nu}$ and 
orbital angular momentum $\hat{l}_{\nu}$ with $\nu=0$ or $\pm 1$. 
The IV $g$ coefficients are given as $g^{\rm (IV)}_l=\frac{g^{\rm (P)}_l-g^{\rm (N)}_l}{2}$ and $g^{\rm (IV)}_s=\frac{g^{\rm (P)}_s-g^{\rm (N)}_s}{2}$ \cite{2011Fujita_M1_GT,1963Kurath,70Eisenberg}, 
where 
$g^{\rm (P)}_l=1$, $g^{\rm (N)}_l=0$, $g^{\rm (P)}_s=5.586$, and $g^{\rm (N)}_s=-3.826$.}
For simplicity, we neglect the quenching effect on $g$ coefficients \cite{2009Vesely,2010Nest,1998VNC}. 
Note that, from the formalism of reduced matrix elements of $\hat{l}_{\nu}$ and $\hat{s}_{\nu}$, the one-body-operator M1 response seldom appears when the SO-partner orbits are both occupied or empty. 
The M1-excitation strength can be obtained as 
\beq
  \frac{dB_{\rm M1}}{d E_{\gamma}} = \sum_{i} \delta(E_{\gamma}-\hbar \omega_i) \sum_{\nu} \abs{\Braket{\omega_i | \oprt{Q}_{\nu}^{\rm (IV)} |\Phi}}^2, \label{eq:BEM}
\eeq
for all the positive QRPA eigenvalues, $\hbar \omega_i >0$.
Note that, in this work, we neglect the effect of the meson-exchange current as well as the couplings to complex configurations \cite{2010Heyde_M1_Rev,1990Richter,2008Marcucci,1994Moraghe,1982Bertsch,2006Ichimura}, which need further multi-body operations going beyond the standard QRPA method.

\begin{figure}[t] \begin{center}
  \includegraphics[width = 0.81\hsize]{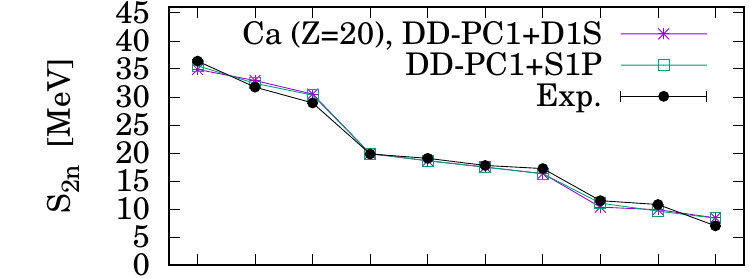}
  \includegraphics[width = 0.81\hsize]{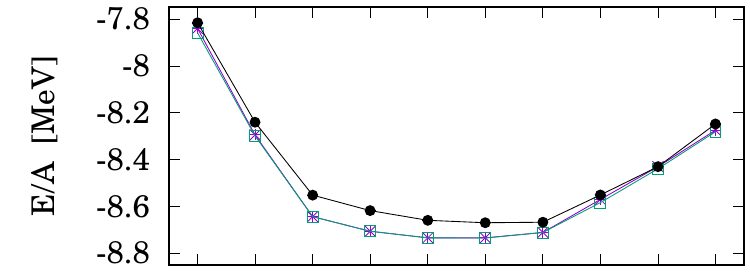}
  \includegraphics[width = 0.81\hsize]{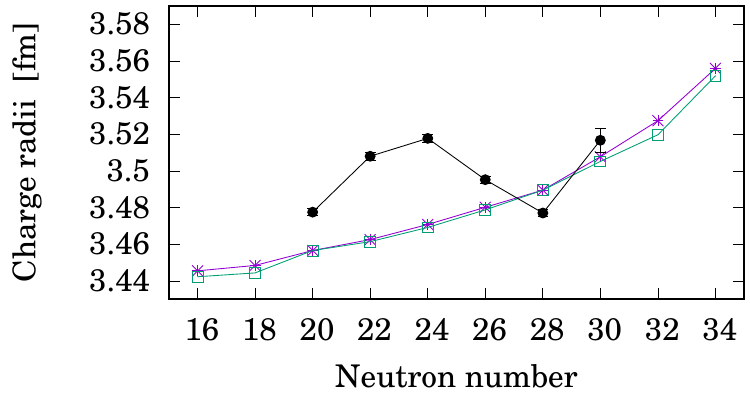}
  \caption{(Top) The two-neutron separation energies for Ca isotopes obtained with our RHB calculation. The experimental data are taken from ref. \cite{NNDCHP}. 
(Middle) The binding energies with the data from ref. \cite{NNDCHP}. 
(Bottom) The charge radii with the data from ref. \cite{2013Angeli}.} \label{fig:0257}
\end{center} \end{figure}
\begin{figure}[t] \begin{center}
  \includegraphics[width = 0.81\hsize]{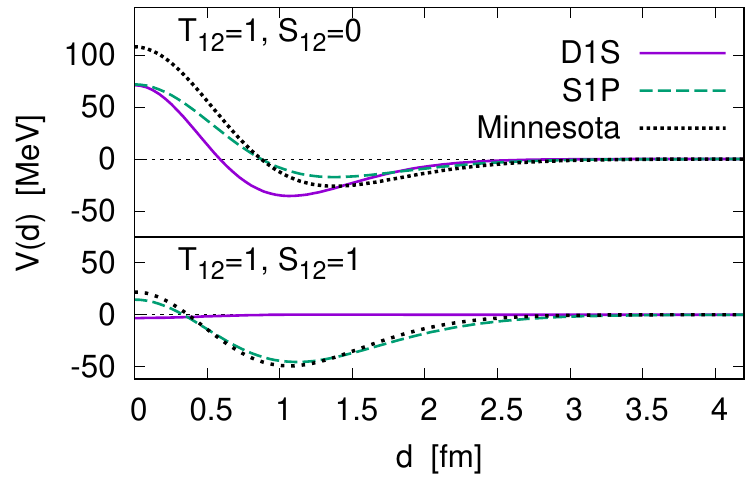}
  \caption{The D1S and S1P-pairing potentials. The Minnesota potential used in ref. \cite{2019OP} is also plotted for comparison.} \label{fig:0256}
\end{center} \end{figure}

For the particle-particle channel, 
with the non-relativistic approximation similarly in ref. \cite{2020Oishi,2005Vret,2003Paar,2008Niksic,2014Niksic}, we employ the Gaussian pairing potential, which was inspired from the Gogny interaction \cite{1980Gogny,1991Berger,2007Hila,2009Goriely,2012Robledo}. 
However, for simplicity,
{\colourbb Gogny zero-range terms like the spin-orbit one are omitted, and only the central part is employed.}
That is, 
\beq
  V_{\rm pp} = \sum_{i=a,b} \left\{ (W_i -H_i) +(B_i-M_i)\hat{P}_{\sigma}  \right\} e^{-\frac{d^2}{\mu^2_i}},
\eeq
where $d$ is the relative distance, $d=\abs{\bir_2-\bir_1}$, and $\hat{P}_{\sigma}$ is the spin-exchange operator.
For its parameters, 
{\colourr the first option in this work is the ``D1S'' pairing interaction \cite{1991Berger}.
The combination of the particle-hole DD-PC1 and the particle-particle D1S parameters well reproduces the empirical ground-state energies and pairing gaps, as shown in the previous RHB calculations \cite{2005Vret,2003Paar,2008Niksic,2014Niksic,2020Oishi}.}
We introduce in parallel a new set of parameters as ``S1P'' (spin-triplet promoting) pairing interaction, in order to check the sensitivity of the M1 response to the pairing modes. 
{\colourr 
The $S_{12}=1$ part of the S1P parameterization is adjusted to mimic the Minnesota potential \cite{77Thom}. 
The Minnesota potential was originally adjusted to the vacuum-scattering parameters of two nucleons, and also has been utilized as an effective-pairing potential \cite{88Suzuki_COSM,03Myo,10Myo,2019OP}. 
}
For the $S_{12}=0$ part of S1P, on the other side, it is tuned to reproduce the same two-neutron separation energy of $^{42}$Ca obtained with the D1S pairing. 
{\colourr Since the $^{42}$Ca nucleus can be well approximated as the doubly-magic $^{40}$Ca core plus two valence neutrons, the gap of binding energies of $^{40}$Ca and $^{42}$Ca, namely the two-neutron separation energy $S_{\rm 2n}$, is a suitable reference for the effective pairing strength.}
The S1P parameters are {\color{black} given as 
$\mu_i=(0.82006,~1.4665)$ fm,
$W_i=(-1144,~68.92)$ MeV,
$B_i=(864.5,-108.716)$ MeV,
$H_i=(-1277,~158.645)$ MeV, and 
$M_i=(864.5,-90.071)$ MeV} for $i=(a,b)$, respectively.
{\colourr 
Note that the present S1P parameterization is just one tool of Gaussian pairing. 
Even if one employs it for the Gogny-HFB calculations \cite{2007Hila,2009Goriely,2012Robledo}, the S1P interaction has no guarantee to reproduce the empirical data. 
Also, the setting of the D1S and S1P pairing models is unique both in the RHB and RQRPA calculations, 
as well as for all the isotopes in the following sections. 
}

\section{Results and Discussions} \label{Sec:III}
First, we verify how the two pairing interactions, D1S and S1P, describe the ground state properties. In Fig. \ref{fig:0257}, the two-neutron separation energies, $S_{\rm 2n}$, the total-binding energies, 
and charge radii obtained with the RHB model are displayed for even-even isotopes $^{36-54}$Ca. 
The DD-PC1 interaction is used in the particle-hole channel. 
One finds that the D1S and S1P coincide well and equivalently reproduce the empirical $S_{\rm 2n}$ values, as well as the GS-binding energies. 
Some deviations to the experimental data are obtained for charge radii in light Ca isotopes, as expected from mean-field calculation \cite{2019Yuksel}. 
{\colourbb We checked that our GS properties are consistent to the non-relativistic Gogny-HFB solutions \cite{2007Hila}.}

The results show that the two pairing models are of the same quality in the description of the GS properties. 
{\colourb 
We have also checked the consistency of the D1S and S1P pairing models in the Sn and Pb isotopes.
The respective results are separately summarized in the Appendix, whereas we focus on the Ca isotopes in the following sections.
}

Figure \ref{fig:0256} shows the actual form of the D1S and S1P pairing potentials. 
Although their GS solutions are of the same quality, their physical properties are different. 
{\colourbb
According to the potential curve in each channel, the D1S potential provides a 
large (small) attractive force in the $S_{12}=0$ ($S_{12}=1$) channel \cite{1991Berger}. 
In contrast, the S1P provides a larger attraction in the $S_{12}=1$ channel. 
Namely, the D1S pairing model mainly supports the SS pairing, 
whereas the S1P may enhance the ST pairing. 
The comparison of the D1S and S1P thus enables us to check the pairing-mode effect. 
}

\begin{figure}[tb] \begin{center}
  \includegraphics[width = 0.9\hsize]{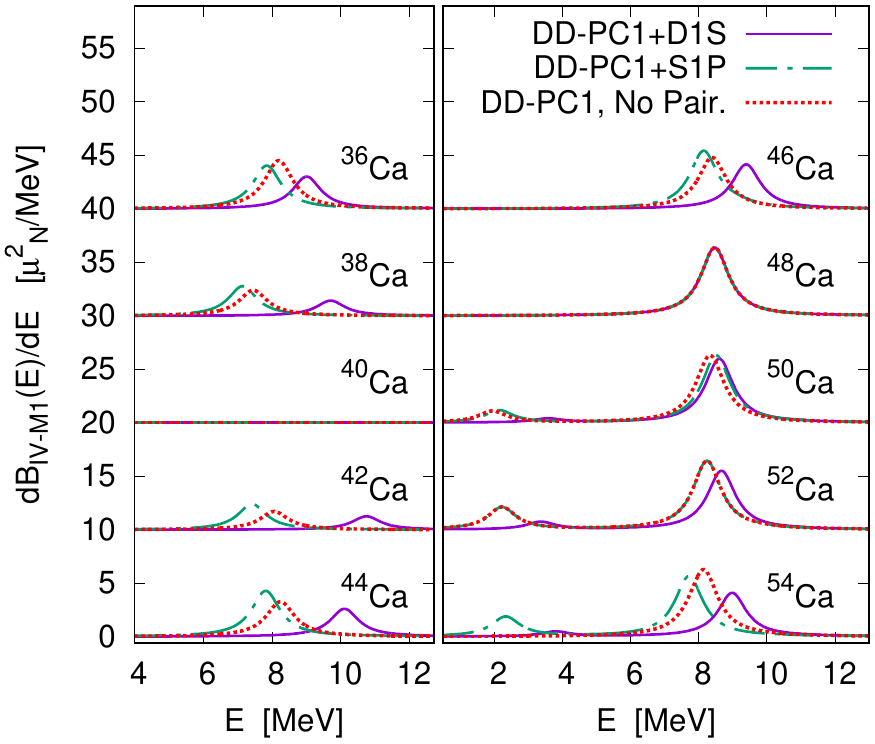}
  \caption{The IV-M1 strength ($0^+_{\rm GS}\longrightarrow 1^+$) in the DD-PC1 plus D1S pairing, plus S1P pairing, and no pairing cases for Ca isotopes. } \label{fig:1123}
\end{center} \end{figure}

\begin{figure}[tb] \begin{center}
  \includegraphics[width = 0.81\hsize]{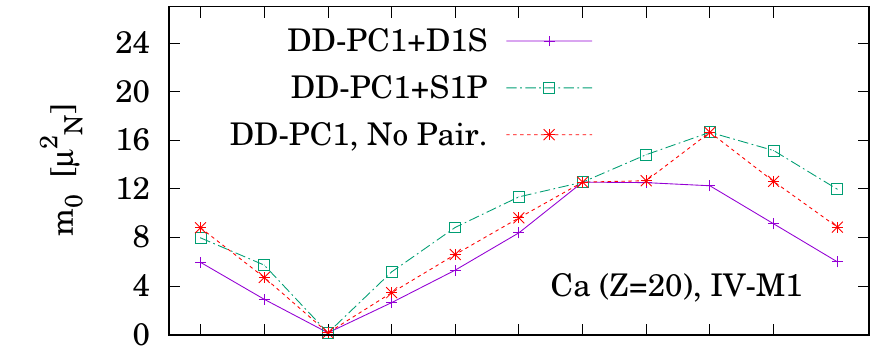}
  \includegraphics[width = 0.81\hsize]{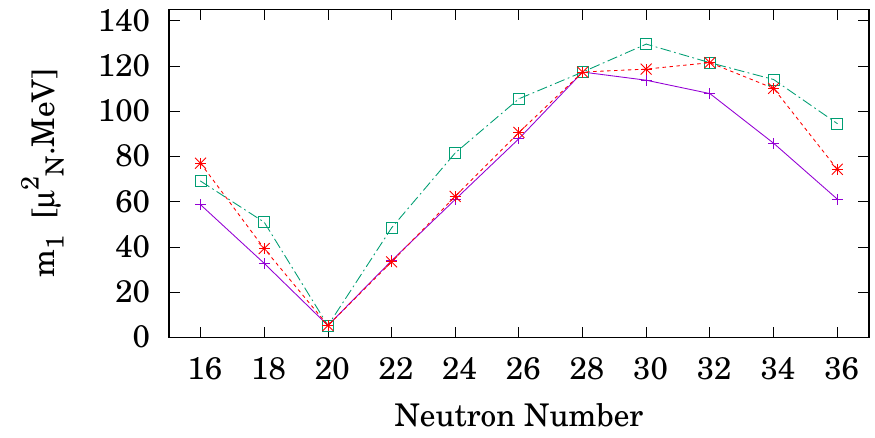}
  \caption{(Top) Non-energy-weighted summation $m_0$ of the M1 strength in the DD-PC1 plus D1S pairing, +S1P pairing, and no pairing cases. (Bottom) Same but for the energy-weighted summation $m_1$.} \label{fig:2452}
\end{center} \end{figure}

In the next step, we explore the sensitivity of the M1 excitation properties on the pairing potential involved.
In Fig. \ref{fig:1123}, the M1-excitation strength of Ca isotopes obtained with the D1S and S1P pairing models are compared. Here the discrete strength in Eq. (\ref{eq:BEM}) is smeared with the Lorentzian using the width of $1.0$ MeV. For comparison, we also present the results by completely neglecting the pairing correlations.
Apparently, the results show a significant difference for open-shell nuclei.
The D1S (S1P) pairing model concludes the higher (lower) M1-excitation energy as well as the reduction (enhancement) of the strength $B_{\rm M1}(E)$ with respect to the no-pairing case.
This model sensitivity can be attributed to the SS and/or ST components controlled by the pairing interaction. 
Since the S1P may support the ST pairing, its M1 response can be enhanced \cite{2019OP}. 
{\colourr Therefore, although the two pairing models provide the same GS energies, 
it does not guarantee the same result for the M1 response.
}

Figure \ref{fig:2452} shows the non-energy-weighted and energy-weighted summations of the M1-strength distributions. That is, $m_{k} \equiv \int E^k \frac{dB_{\rm M1}}{dE} dE$, 
where $k=0$ and $1$.
Consistently to the strength distribution in Fig. \ref{fig:1123}, the D1S (S1P) model provides the reduction (enhancement) of $m_0$ values, consistently to ref. \cite{2019OP,2020Oishi}. 
{\colourb 
The $m_0$ ($m_1$) values obtained with the D1S and S1P pairing models show the $20$-$40$\% ($10$-$30$\%) difference for open-shell Ca nuclei. 
The systematic measurement of M1-excitation strength with the corresponding accuracy may enable one to qualify whether the SS and/or ST pairing models are appropriate or not. 
}

{\colourr 
There is a connection between the $m_k$ values and the GS structure. 
In Fig. \ref{fig:2452}, for $N=22$-$26$ as example, the S1P results show higher $m_k$ values than the D1S and no-pairing cases. 
In parallel, we checked that there is a finite contribution of the $2p_{3/2}$ orbits in addition to the $1f_{7/2}$ in these RHB-ground states, due to the pairing attraction in the ST channel. 
In contrast, the D1S pairing force cannot invoke this mixture, and its GS solution is almost of $1f_{7/2}$ similarly to the no-pairing case.
Then as expected, their $m_k$ values are similar in Fig. \ref{fig:2452}. 
For the other Ca isotopes, the similar connection exists, but with the different orbit(s) involved or not depending on the chosen pairing model. 
Note that our total $m_0({\rm M1})$ values are consistent with the Gogny-QRPA results \cite{2016Goriely}. 
}

\begin{figure}[tb] \begin{center}
  \includegraphics[width = 0.9\hsize]{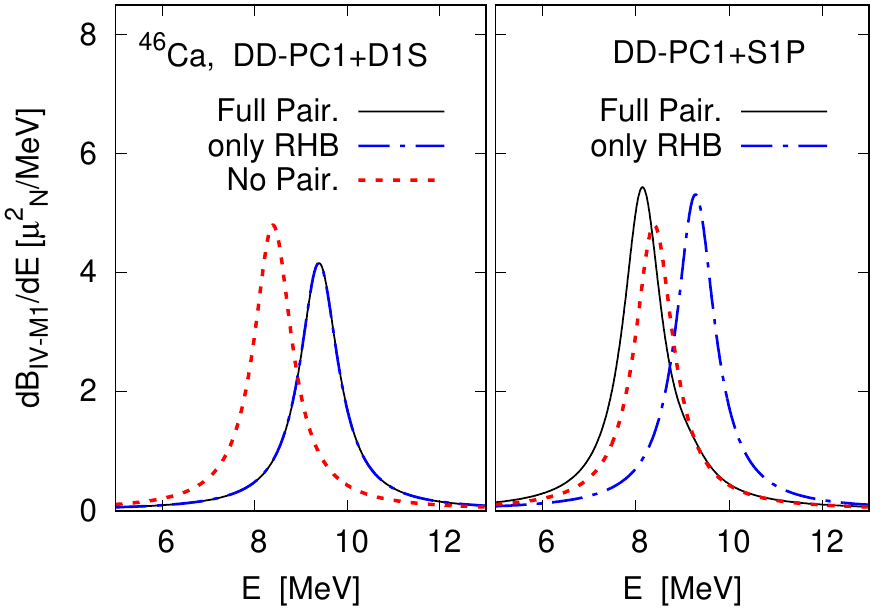}
  \caption{(Left panel) The IV-M1 strength for $^{46}$Ca by using the D1S-pairing interaction. The ``Full Pair.'' result (solid line) indicates that the pairing interaction is used in both the RHB and RQRPA, whereas, in the ``only RHB'' (dot-broken line), it is neglected from the RQRPA residual interaction. The no-pairing result is also plotted. (Right panel) The same but obtained with the S1P-pairing interaction.} \label{fig:JA4262}
\end{center} \end{figure}

We mention the generalized rule of energy-weighted summation by Hinohara \cite{2019Hinohara_SR}.
As a substantial conclusion by Hinohara, the energy-weighted summation within the framework of arbitrary energy-density functional (EDF) can be independent of the pairing interaction. The conditions to hold this conclusion include that (i) the chosen EDF keeps the local-gauge invariance, (ii) the excitation operator is of the coordinate type, and (iii) there is no momentum-dependent ingredient.
In contrast, the current pairing interactions of D1S and S1P have the finite range, which can be related with the momentum-dependence in the particle-particle channel. 
Also, because the M1 operator is spin-dependent, the present calculation may not simply apply to the Hinohara's case \cite{2019Hinohara_SR}. 
As the result, the present energy-weighted summation with the D1S or S1P interaction may differ from the no-pairing one.

\begin{figure}[tb] \begin{center}
  \includegraphics[width = 0.9\hsize]{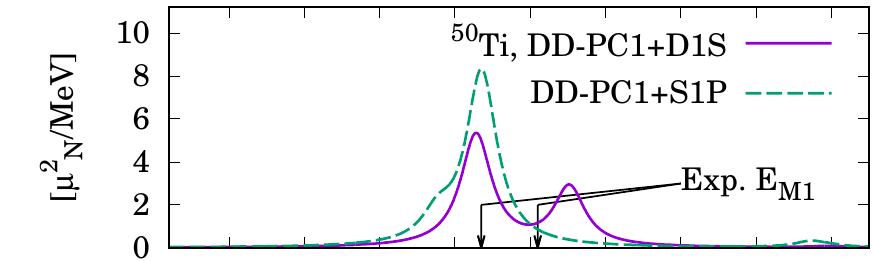}
  \includegraphics[width = 0.9\hsize]{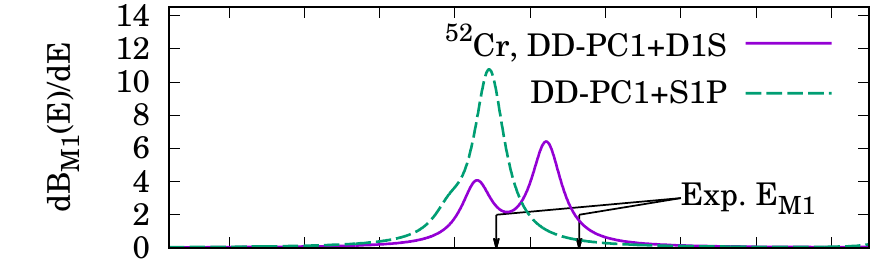}
  \includegraphics[width = 0.9\hsize]{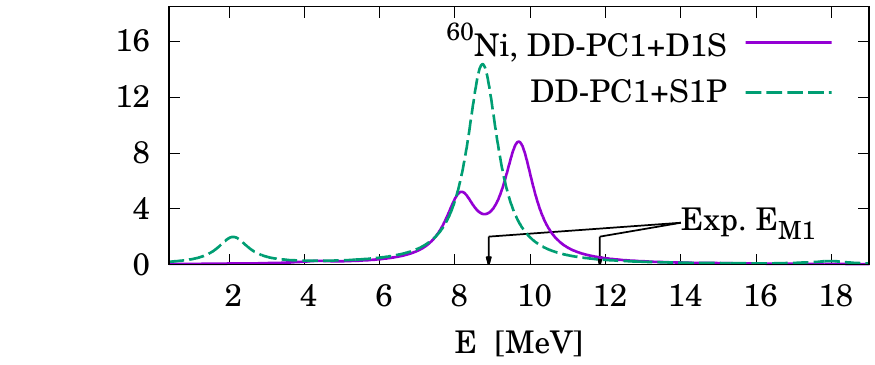}
  \caption{The IV-M1 strength ($0^+_{\rm GS}\longrightarrow 1^+$) for $^{50}$Ti, $^{52}$Cr, and $^{60}$Ni by using the D1S and S1P pairing interactions. The two major M1-excitation energies in each nucleus observed in the experiments are also shown \cite{1983Crawley,1985Sober}.} \label{fig:FW8644}
\end{center} \end{figure}

{\colourr 
In Fig. \ref{fig:JA4262}, for $^{46}$Ca as an example, we compare the two cases, where the pairing interactions are included or neglected in the RQRPA calculations.
This test clarifies the pairing effect of the two models in the RQRPA channel. 
Note that the IV-PV interaction is still active in the the RQRPA channel. 
In the left panel of Fig. \ref{fig:JA4262}, with the D1S pairing, the ``Full Pair.'' result coincides with the 
``only RHB'' one.
Thus, it is shown that the D1S pairing does not contribute effectively in the RQRPA residual interactions: 
this result was confirmed also in ref. \cite{2020Oishi}. 
In contrast, in the right panel of Fig. \ref{fig:JA4262}, 
the M1-excitation energy noticeably increases by omitting the S1P pairing interaction in the RQRPA. 
Namely, in the RQRPA for $1^+$ configuration, the S1P-pairing interaction works as the strong attraction, whereas the D1S does not. 
}
This result can be understood from their matrix elements, i.e., only the ST force can provide the finite matrix elements in the $1^+$ configuration.
For more details, see ref. \cite{2019OP,2012Tani}.

Figure \ref{fig:FW8644} shows our results of several open-shell nuclei, for which the experimental data are available \cite{1983Crawley,1985Sober}.
For $^{50}$Ti as example, in the RHB+RQRPA calculation with the D1S pairing interactions, there are neutron-M1 and proton-M1 peaks at $8.6$ and $11$ MeV, respectively. 
This result qualitatively agree with the two-peak distribution observed in the experiment \cite{1985Sober}. By analyzing the RQRPA transition components, we found that these two peaks can be mostly attributed to the $1f_{7/2} \longrightarrow 1f_{5/2}$ transition of the valence protons and neutrons above the $^{40}$Ca core. 
In the S1P case, on the other side, the result becomes rather different than the experimental data. This discrepancy is linked with the effect of the RQRPA-residual interaction shown in Fig. \ref{fig:JA4262}. Because the S1P interaction decreases the M1-excitation energy, the neutron-M1 and proton-M1 transitions of $^{50}$Ti have peaks at $8.7$ and $7.5$ MeV, respectively.
The similar behaviour can be found in the other nuclei, $^{52}$Cr and $^{60}$Ni, where only the D1S pairing interaction can reproduce the fragmented distribution.
By comparing with the experimental data, the D1S model supporting the SS-pairing mode provides a better option for the description of M1-transition properties up to the one-body QRPA level.
{\colourb 
Simultaneously, the strong attractive pairing in the ST channel has not been validated in the present study.
}

{\colourr 
Notice that the pairing-model sensitivity shown in the present results is independent of the M1-quenching effect. 
The quenching factors on $g_s$ and $g_l$ may change the $B({\rm M1})$ height by the constant ratio, but commonly for the D1S and S1P-pairing cases. 
Namely, even if we employ these quenching factors, the difference between the D1S and S1P-pairing results cannot be cancelled.
}

\section{Summary} \label{Sec:summary}
Summarizing this work, we have investigated the pairing-mode sensitivity of the IV-M1 excitation. 
Within the RNEDF framework, the relativistic Hartree-Bogoliubov model and relativistic QRPA are employed, using the density-dependent point-coupling interaction. 
The effects of the spin-singlet and spin-triplet pairing channels have been considered. 
It is shown that the M1-excitation energy and strength display a sizable sensitivity to the spin-singlet and spin-triplet pairing interactions involved, while at the ground-state level they provide the results with the same quality. 
{\colourb 
The strong attractive pairing in the ST channel has not been validated in the present study.
}
The accurate measurement of the M1 excitation in open-shell nuclei can provide further information to determine the effect of the SS and ST-pairing modes inside finite nuclei. 
In parallel, we note that further developments of calculation are on demand to account for the effects of nuclear deformation as well as the couplings to higher-order configurations for the M1 excitation. 
{\colourbb
Even though it is independent of our main conclusion and thus neglected in this work, 
checking the M1-quenching effect on the observables is also one future task, 
including the modification of gyromagnetic factors as $g\longrightarrow \eta g$. 
}

Sensitivity of the other observables than the M1 response may provide complementary information on the pairing modes inside nuclei.
Those include, e.g., the Gamow-Teller transitions, electric-multipole excitations, moment of inertia with deformation, etc. 
We aim to account for these topics in forthcoming studies.

\begin{acknowledgements}
We especially thank Tamara Nik\v{s}i\'{c} and Dario Vretenar for fruitful discussions. 
This work is supported by  the ``QuantiXLie Centre of Excellence'', a project co-financed by the Croatian Government and European Union through the European Regional Development Fund, the Competitiveness and Cohesion Operational Programme (KK.01.1.1.01). 
\end{acknowledgements}

\begin{figure}[t] \begin{center}
  \includegraphics[width = 0.95\hsize]{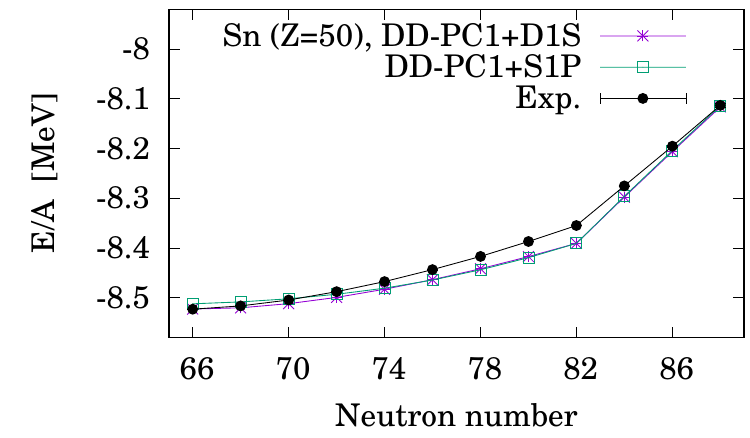}
  \includegraphics[width = 0.95\hsize]{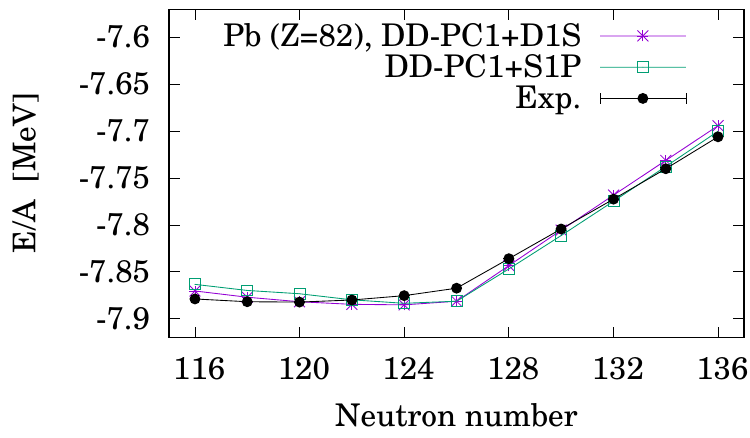}
  \caption{Binding energies for Sn and Pb isotopes obtained with the DD-PC1+D1S and DD-PC1+S1P interactions. The experimental data are taken from ref. \cite{NNDCHP}. 
} \label{fig:CUTFER}
\end{center} \end{figure}

{\colourb 
\appendix
\section{Comparison of pairing models}
In the main text, we have utilized the D1S and S1P-pairing models combined with the relativistic DD-PC1 interaction for the RHB calculations. 
In this Appendix, we check the consistency between D1S and S1P in terms of the binding energies for open-shell nuclei. 
We perform the RHB calculations for the Sn and Pb isotopes by keeping the same setting and parameters used in the main text for Ca isotopes.

In Fig. \ref{fig:CUTFER}, our RHB results for Sn and Pb isotopes are displayed. 
The D1S and S1P-pairing models show a good agreement in terms of the binding energies per nucleon, $E/A$. 
Namely, they have the same quality on the reproduction of the ground-state energies of open-shell nuclei. 
Notice also that their results exactly coincide for the doubly-magic nuclei. 
The error from the experimental $E/A$ data is, even at maximum, less than $0.1$ MeV throughout the Ca, Sn, and Pb isotopes. 
We have also checked that the RHB results, which include the pairing strength, quasiparticle spectrum, and charge radii, are well equivalent between the DD-PC1+D1S and DD-PC1+S1P cases.
}



\begin{thebibliography}{10}
\providecommand{\url}[1]{{#1}}
\providecommand{\urlprefix}{URL }
\expandafter\ifx\csname urlstyle\endcsname\relax
  \providecommand{\doi}[1]{DOI \discretionary{}{}{}#1}\else
  \providecommand{\doi}{DOI \discretionary{}{}{}\begingroup
  \urlstyle{rm}\Url}\fi

\bibitem{57Bardeen}
J.~Bardeen, L.N. Cooper, J.R. Schrieffer, Phys. Rev. \textbf{108}, 1175 (1957)

\bibitem{1972Osheroff_01}
D.D. Osheroff, R.C. Richardson, D.M. Lee, Phys. Rev. Lett. \textbf{28}, 885
  (1972)

\bibitem{1972Osheroff_02}
D.D. Osheroff, W.J. Gully, R.C. Richardson, D.M. Lee, Phys. Rev. Lett.
  \textbf{29}, 920 (1972)

\bibitem{1972Leggett}
A.J. Leggett, Phys. Rev. Lett. \textbf{29}, 1227 (1972)

\bibitem{03Dean_rev}
D.J. Dean, M.~Hjorth-Jensen, Rev. Mod. Phys. \textbf{75}, 607 (2003)

\bibitem{05BB}
D.~Brink, R.~Broglia, \emph{Nuclear Superfluidity: Pairing in Finite Systems}.
\newblock Cambridge Monographs on Particle Physics, Nuclear Physics and
  Cosmology (Cambridge University Press, Cambridge, UK, 2005)

\bibitem{13BZ}
R.A. Broglia, V.~Zelevinsky (eds.), \emph{Fifty Years of Nuclear BCS: Pairing
  in Finite Systems} (World Scientific, Singapore, 2013)

\bibitem{96Jacek}
J.~Dobaczewski, W.~Nazarewicz, T.R. Werner, J.F. Berger, C.R. Chinn,
  J.~Decharg\'e, Phys. Rev. C \textbf{53}, 2809 (1996)

\bibitem{04Suzu}
Y.~Suzuki, H.~Matsumura, B.~Abu-Ibrahim, Phys. Rev. C \textbf{70}, 051302
  (2004)

\bibitem{2005Sand}
N.~Sandulescu, P.~Schuck, X.~Vi\~nas, Phys. Rev. C \textbf{71}, 054303 (2005)

\bibitem{08Hagi}
K.~Hagino, N.~Takahashi, H.~Sagawa, Phys. Rev. C \textbf{77}, 054317 (2008)

\bibitem{2008Bulgac}
A.~Bulgac, Y.~Yu, Phys. Rev. Lett. \textbf{88}, 042504 (2002)

\bibitem{2013Pastore}
A.~Pastore, J.~Margueron, P.~Schuck, X.~Vi\~nas, Phys. Rev. C \textbf{88},
  034314 (2013)

\bibitem{1968Tamagaki}
R.~Tamagaki, Progress of Theoretical Physics \textbf{39}(1), 91 (1968)

\bibitem{1970Tamagaki}
R.~Tamagaki, Progress of Theoretical Physics \textbf{44}(4), 905 (1970)

\bibitem{1980Gogny}
J.~Decharg\'e, D.~Gogny, Phys. Rev. C \textbf{21}, 1568 (1980)

\bibitem{1991Berger}
J.F. Berger, M.~Girod, D.~Gogny, Computer Physics Communications \textbf{63},
  365 (1991)

\bibitem{88Suzuki_COSM}
Y.~Suzuki, K.~Ikeda, Phys. Rev. C \textbf{38}, 410 (1988)

\bibitem{03Myo}
T.~Myo, S.~Aoyama, K.~Kat\ifmmode~\bar{o}\else \={o}\fi{}, K.~Ikeda, Physics
  Letters B \textbf{576}(3-4), 281  (2003)

\bibitem{10Myo}
T.~Myo, R.~Ando, K.~Kat\ifmmode~\bar{o}\else \={o}\fi{}, Physics Letters B
  \textbf{691}(3), 150  (2010)

\bibitem{2014Oishi}
T.~Oishi, K.~Hagino, H.~Sagawa, Phys. Rev. C \textbf{90}, 034303 (2014)

\bibitem{2019OP}
T.~Oishi, N.~Paar, Phys. Rev. C \textbf{100}, 024308 (2019)

\bibitem{2020Oishi}
T.~Oishi, G.~Kru\v{z}i\'{c}, N.~Paar, Journal of Physics G: Nuclear and
  Particle Physics \textbf{47}, 115106 (2020)

\bibitem{1985Rich}
A.~Richter, Progress in Particle and Nuclear Physics \textbf{13}, 1  (1985)

\bibitem{1990Rich}
A.~Richter, Nuclear Physics A \textbf{507}(1), 99  (1990)

\bibitem{2010Heyde_M1_Rev}
K.~Heyde, P.~von Neumann-Cosel, A.~Richter, Rev. Mod. Phys. \textbf{82}, 2365
  (2010), and references therein.

\bibitem{2008Pietralla_Rev}
N.~Pietralla, P.~von Brentano, A.~Lisetskiy, Progress in Particle and Nuclear
  Physics \textbf{60}(1), 225  (2008)

\bibitem{2011Fujita_M1_GT}
Y.~Fujita, B.~Rubio, W.~Gelletly, Progress in Particle and Nuclear Physics
  \textbf{66}(3), 549  (2011)

\bibitem{2006Speth}
F.~Gr\"{u}mmer, J.~Speth, Journal of Physics G: Nuclear and Particle Physics
  \textbf{32}(7), R193 (2006)

\bibitem{2009Vesely}
P.~Vesely, J.~Kvasil, V.O. Nesterenko, W.~Kleinig, P.G. Reinhard, V.Y.
  Ponomarev, Phys. Rev. C \textbf{80}, 031302(R) (2009)

\bibitem{2010Nest}
V.O. Nesterenko, J.~Kvasil, P.~Vesely, W.~Kleinig, P.G. Reinhard, V.Y.
  Ponomarev, Journal of Physics G: Nuclear and Particle Physics \textbf{37}(6),
  064034 (2010)

\bibitem{2010Nest_2}
V.O. Nesterenko, J.~Kvasil, P.~Vesely, W.~Kleinig, P.G. Reinhard, Int. J. Mod.
  Phys. E \textbf{19}, 558 (2010)

\bibitem{2019Tselyaev}
V.~Tselyaev, N.~Lyutorovich, J.~Speth, P.G. Reinhard, D.~Smirnov, Phys. Rev. C
  \textbf{99}, 064329 (2019)

\bibitem{2020Speth}
J.~Speth, P.G. Reinhard, V.~Tselyaev, N.~Lyutorovich, preprint available: p.
  arXiv: 2001.07236 (2020)

\bibitem{2020Kruzic}
G.~Kru\ifmmode \check{z}\else \v{z}\fi{}i\ifmmode~\acute{c}\else \'{c}\fi{},
  T.~Oishi, D.~Vale, N.~Paar, Phys. Rev. C \textbf{102}, 044315 (2020)

\bibitem{1974Walecka}
J.D. Walecka, Annals of Physics \textbf{83}, 491 (1974)

\bibitem{1977Boguta}
J.~Boguta, A.~Bodmer, Nuclear Physics A \textbf{292}, 413 (1977)

\bibitem{1989Reinhard}
P.G. Reinhard, Reports on Progress in Physics \textbf{52}, 439 (1989)

\bibitem{2005Vret}
D.~Vretenar, A.V. Afanasjev, G.A. Lalazissis, P.~Ring, Physics Report
  \textbf{409}, 101 (2005), and references therein.

\bibitem{2006Meng}
J.~Meng, H.~Toki, S.~Zhou, S.~Zhang, W.~Long, L.~Geng, Progress in Particle and
  Nuclear Physics \textbf{57}(2), 470  (2006)

\bibitem{2016Ebran_PRC}
J.P. Ebran, A.~Mutschler, E.~Khan, D.~Vretenar, Phys. Rev. C \textbf{94},
  024304 (2016)

\bibitem{2003Paar}
N.~Paar, P.~Ring, T.~Nik\v{s}i\'{c}, D.~Vretenar, Phys. Rev. C \textbf{67},
  034312 (2003)

\bibitem{2008Niksic}
T.~Nik\ifmmode \check{s}\else \v{s}\fi{}i\ifmmode~\acute{c}\else \'{c}\fi{},
  D.~Vretenar, P.~Ring, Phys. Rev. C \textbf{78}, 034318 (2008)

\bibitem{2014Niksic}
T.~Nik\v{s}i\'{c}, N.~Paar, D.~Vretenar, P.~Ring, Computer Physics
  Communications \textbf{185}(6), 1808  (2014)

\bibitem{1963Kurath}
D.~Kurath, Phys. Rev. \textbf{130}, 1525 (1963)

\bibitem{70Eisenberg}
J.~Eisenber, W.~Greiner, \emph{Nuclear Theory Volume 2: Excitation Mechanisms
  of the Nucleus} (North-Holland Publishing Company, Amsterdam, 1970)

\bibitem{1998VNC}
P.~von Neumann-Cosel, A.~Poves, J.~Retamosa, A.~Richter, Physics Letters B
  \textbf{443}(1), 1  (1998)

\bibitem{1990Richter}
A.~Richter, A.~Weiss, O.~Ha\"usser, B.A. Brown, Phys. Rev. Lett. \textbf{65},
  2519 (1990)

\bibitem{2008Marcucci}
L.E. Marcucci, M.~Pervin, S.C. Pieper, R.~Schiavilla, R.B. Wiringa, Phys. Rev.
  C \textbf{78}, 065501 (2008)

\bibitem{1994Moraghe}
S.~Moraghe, J.~Amaro, C.~García-Recio, A.~Lallena, Nuclear Physics A
  \textbf{576}(4), 553  (1994)

\bibitem{1982Bertsch}
G.F. Bertsch, I.~Hamamoto, Phys. Rev. C \textbf{26}, 1323 (1982)

\bibitem{2006Ichimura}
M.~Ichimura, H.~Sakai, T.~Wakasa, Progress in Particle and Nuclear Physics
  \textbf{56}(2), 446  (2006)

\bibitem{NNDCHP}
\urlprefix\url{http://www.nndc.bnl.gov/chart/}.
\newblock ``Chart of Nuclides'', National Nuclear Data Center (NNDC);
  http://www.nndc.bnl.gov/chart/

\bibitem{2013Angeli}
I.~Angeli, K.~Marinova, Atomic Data and Nuclear Data Tables \textbf{99}(1), 69
  (2013)

\bibitem{2007Hila}
S.~Hilaire, M.~Girod, The European Physical Journal A \textbf{33}, 237 (2007),
  with data at [http://www-phynu.cea.fr/ science\_en\_ligne/
  carte\_potentiels\_microscopiques/ carte\_potentiel\_nucleaire\_eng.htm]

\bibitem{2009Goriely}
S.~Goriely, S.~Hilaire, M.~Girod, S.~P\'eru, Phys. Rev. Lett. \textbf{102},
  242501 (2009)

\bibitem{2012Robledo}
L.M. Robledo, R.~Bernard, G.F. Bertsch, Phys. Rev. C \textbf{86}, 064313 (2012)

\bibitem{77Thom}
D.~Thompson, M.~Lemere, Y.~Tang, Nuclear Physics A \textbf{286}(1), 53  (1977)

\bibitem{2019Yuksel}
E.~Y\"uksel, T.~Marketin, N.~Paar, Phys. Rev. C \textbf{99}, 034318 (2019)

\bibitem{2016Goriely}
S.~Goriely, S.~Hilaire, S.~P\'eru, M.~Martini, I.~Deloncle, F.~Lechaftois,
  Phys. Rev. C \textbf{94}, 044306 (2016)

\bibitem{2019Hinohara_SR}
N.~Hinohara, Phys. Rev. C \textbf{100}, 024310 (2019)

\bibitem{1983Crawley}
G.~Crawley, C.~Djalali, N.~Marty, Report of Inst. de Physique Nucleaire
  \textbf{15}, IPNO DRE 83 15 (1983)

\bibitem{1985Sober}
D.I. Sober, B.C. Metsch, W.~Kn\"upfer, G.~Eulenberg, G.~K\"uchler, A.~Richter,
  E.~Spamer, W.~Steffen, Phys. Rev. C \textbf{31}, 2054 (1985)

\bibitem{2012Tani}
Y.~Tanimura, K.~Hagino, H.~Sagawa, Phys. Rev. C \textbf{86}, 044331 (2012)

\end{thebibliography}

\end{document}